\documentclass[prl,twocolumn,aps,amsmath,amssymb]{revtex4-1}
\usepackage{bm}
\usepackage{graphicx}
\usepackage{amsmath}
\usepackage{amssymb}
\usepackage[utf8]{inputenc}
\usepackage[T1]{fontenc}
\usepackage{color}
\usepackage{float}
\usepackage{upgreek}
\usepackage{subfigure}
\usepackage[unicode=true,colorlinks=true,citecolor=blue]{hyperref}
\usepackage{placeins}
\usepackage{lipsum}
\usepackage{epsfig}
\usepackage[normalem]{ulem}
\usepackage{wrapfig}
\usepackage{physics}

\newcommand{\nix}[1]{}
\begin{document}

\title{Highly superlinear photogalvanic effects in (Bi$_{0.3}$Sb$_{0.7}$)$_2$(Te$_{0.1}$Se$_{0.9}$)$_3$: Probing 3D topological insulator surface states at room temperature 
		\newline
}

\author{S.~N.~Danilov$^1$,  L.~E. Golub$^2$, T.~Mayer$^1$, A.~Beer$^1$, S.~Binder$^1$, E.~M\"onch$^1$, J. Minar$^3$, M.~Kronseder$^1$, C. H. Back$^4$, D.~Bougeard$^1$ and S.~D. Ganichev$^{1,5}$}

\affiliation{$^1$Terahertz Center, University of Regensburg, 93040 Regensburg, Germany}
\affiliation{$^2$Ioffe Institute, 194021	St. Petersburg, Russia}
\affiliation{$^3$ New Technologies-Research Center, University of West Bohemia, 301 00 	Plzeň 3, Czech Republic}
\affiliation{$^4$ Technical University Munich, 85748 Garching, Germany}
\affiliation{$^5$ CENTERA, Institute of High Pressure Physics PAS, 01142 Warsaw, Poland}

\begin{abstract}
We report on the observation of complex nonlinear intensity dependence of the circular and linear photogalvanic currents induced by infrared radiation in compensated (Bi$_{0.3}$Sb$_{0.7}$)$_2$(Te$_{0.1}$Se$_{0.9}$)$_3$ 3D topological insulators. 
The photocurrents are induced by direct optical transitions between topological surface and bulk states.
We show that an increase of the radiation intensity results first in a highly superlinear raise of the amplitude of both types of photocurrents,
whereas at higher intensities  the photocurrent saturates. Our analysis of the observed nonlinearities shows that the superlinear behavior of the photocurrents is caused by a heating of the electron gas, while the saturation is induced by a slow relaxation of the  photoexcited carriers resulting in absorbance bleaching. 
The observed nonlinearities give access to the
Fermi level position with respect to the Dirac point and the energy relaxation times of Dirac fermions providing an experimental room temperature probe for topological surface states.
\end{abstract}

\maketitle

\section{Introduction}


Topological insulators (TI) with low dimensional surface states described by the massless Dirac equation have recently moved into the focus of modern research. TIs challenge fundamental physical concepts as well as hold a great potential for applications~\cite{Hasan2010,Moore2010,Qi2011,Ortmann2015,Vanderbilt2018}. Recently photocurrents excited by laser radiation in TI systems attracted growing attention because twithheir study opens up new opportunities for probing topological behavior of Dirac fermions (DF), for review see Ref.~\cite{Dyakonov2017}. In particular, photogalvanic effects (PGE)~\cite{Sturman1992,GanichevPrettl2003,Ivchenko2005} in TIs, including the linear~\cite{Olbrich2014,Kastl2015,Plank2016drag,Braun2016,Kuroda2017,Plank2018,Plank2018_2,Wang2019} and circular~\cite{Hosur2011,McIver2011,Junck2013,Artemenko2013,Kastl2015,Entin2016,Okada2016,Hamh2016,Dantscher2017,Kuroda2017,Pan2017,Plank2018_2,Xu2019,Yu2019,Durnev2019,Wang2019,Meyer2020} PGE provide important information on electronic and spin properties of DF and are thus highly suitable for the study of topological phenomena in various classes of TI materials~\cite{Plank2016,Dantscher2015}, for review see Ref.~\cite{Plank2018_2}. An important advantage of the PGE phenomena is that in most cases they can be used to selectively probe the surface states of TIs even at room temperature~\cite{Dyakonov2017,Plank2016,Dantscher2015}.  The system properties that have experimentally been probed by PGE are manifold: the Fermi velocity, the cyclotron masses as a function of carrier density and temperature, the orientation of surface domains in 3D TIs, and the surface state mobility. A possibility to selectively probe the topological surface states (TSS) is particularly helpful in the search for novel 3D TIs in which electronic transport  experiments are often handicapped by a large residual bulk charge carrier density. In general, PGEs are caused by the redistribution of charge carriers in momentum space induced by incident radiation. The PGE current scales with the second power of the radiation electric field, i.e. linearly with radiation intensity $I$~\cite{Sturman1992,GanichevPrettl2003,Ivchenko2005}. Deviations from this linearity have been observed in quantum wells~\cite{Ganichev2002,Schneider2004} and in graphene~\cite{Candussio2021} and their analysis opened up new opportunities for materials characterization. In TIs, however, the presence of such effects has neither experimentally nor theoretically been addressed.

Here we experimentally demonstrate and theoretically analyze a highly nonlinear intensity dependence of both, linear and circular, photogalvanic effects. We demonstrate that the excitation of  (Bi$_{1-x}$, Sb$_x$)$_2$(Te$_{1-y}$, Se$_y$)$_3$ (BSTS) with high power mid-infrared radiation results in PGE currents, which at moderate intensities exhibit a superlinear intensity dependence, followed by saturation at high $I$. We show that the superlinear dependence results from a radiation-induced electron gas heating which induces a change in the population of initial and final states of direct optical transitions involving the topological surface states. The signal amplitude and its nonlinearity are found to be very sensitive to the position of the Fermi level. Hence, the PGE represents a sensitive room temperature probe for this important parameter. Furthermore,  the photocurrent saturation, observed at very high intensities of hundreds of kW/cm$^2$, is shown to be caused by energy relaxation of photoexcited carriers and, consequently, yields information on DF energy relaxation times. 



\section{Samples and methods}
\label{samples_methods}

\subsection{Samples}
\label{samples}

\begin{figure}[t]
	\centering
	\includegraphics[width=0.8\linewidth]{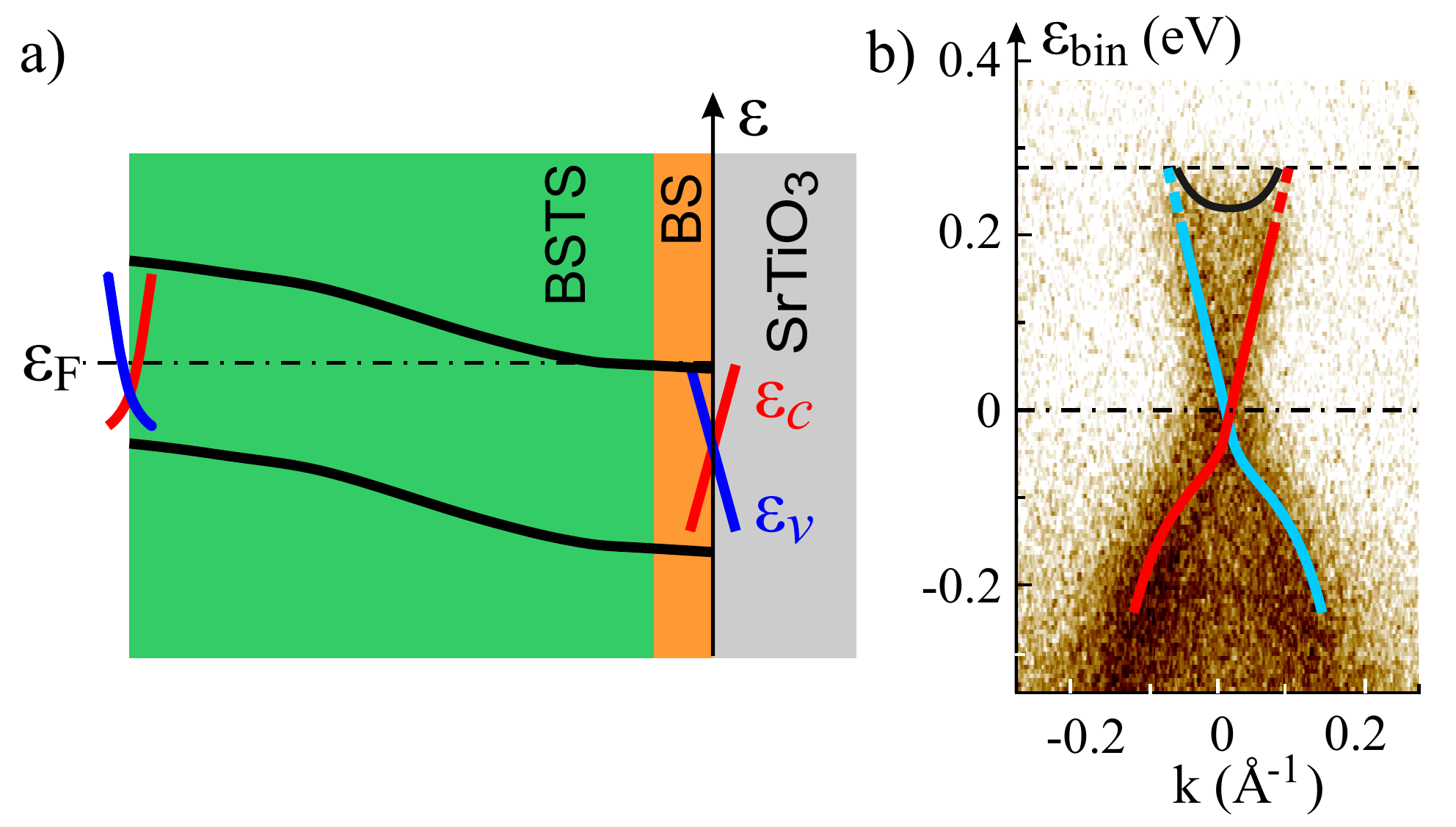}
	\caption{
Band edge scheme of the investigated heterostucture stack. (a) The stack is built from $n$-type Bi$_2$Se$_3$ and a slightly $p$-type (Bi$_{1-x}$Sb$_x$)$_2$(Te$_{1-x}$Se$_x$)$_3$ (($x$|$y$)$\sim$(0.7|0.9)) layer with different thicknesses: Sample \#A: 1 QL BS + 10 QL BSTS, Sample \#B and \#C: 2 QL BS + 20 QL BSTS. This $pn$-heterostructure leads to a band bending effect in growth direction. In our samples~\cite{Mayer2021} the Fermi energy $\varepsilon_F$ evolves from the CBM at the substrate interface (SrTiO$_3$) to a mid-gap position at the top surface ($\varepsilon_c$ refers to the CBM, $\varepsilon_v$ to the VBM). 
%
(b) ARPES at 77K, illustrating the linear dispersion at the top surface of a heterostructure, here, for a sample with the same stoichiometry ($x$|$y$)$\sim$(0.7|0.9), 1 QL BS and 5.6 QL BSTS, after \cite{Mayer2021}.   
}
	\label{fig_01}
\end{figure}

\begin{figure}[t]
	\centering
	\includegraphics[width=1.0\linewidth]{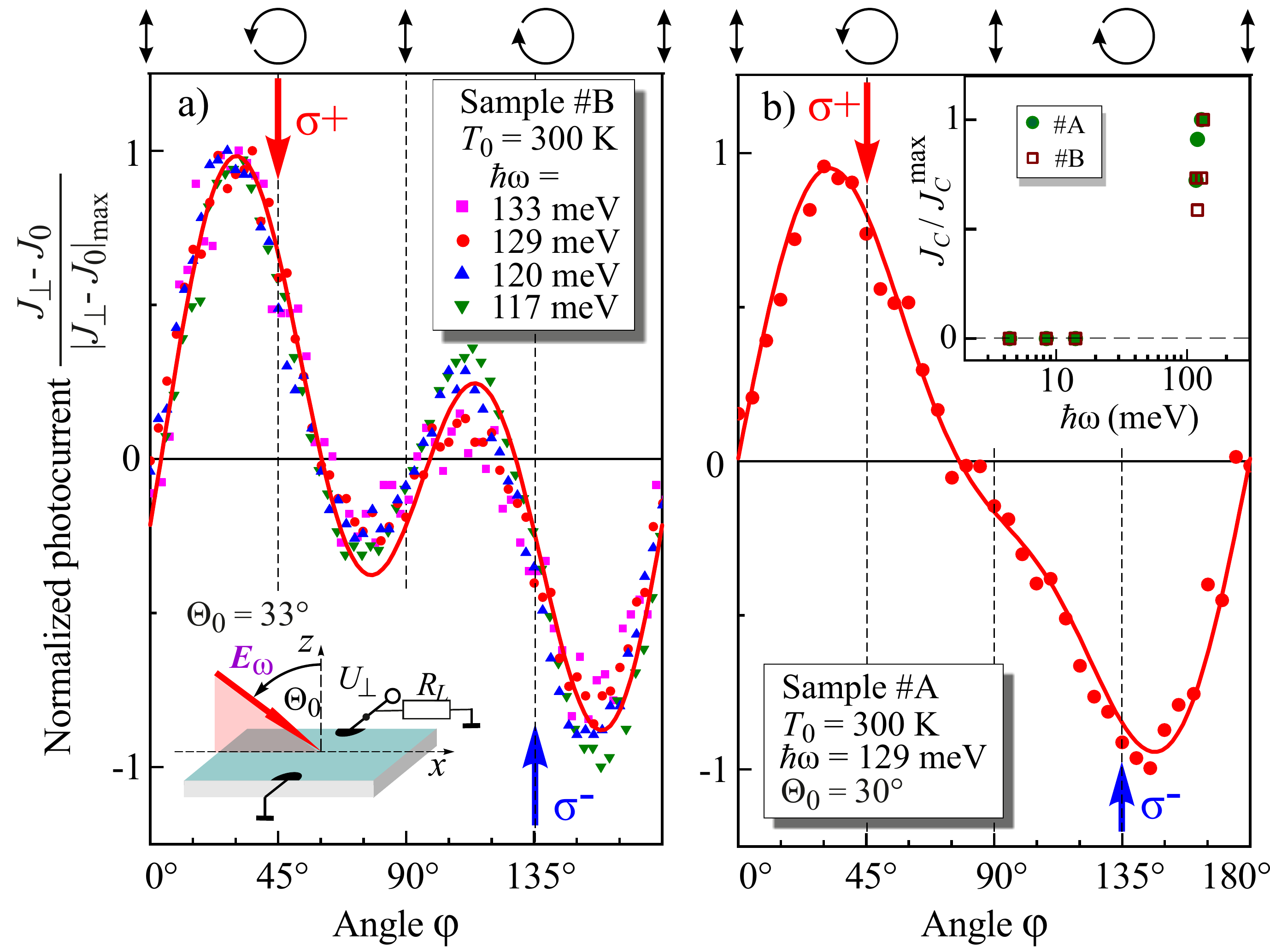}
	\caption{
Transverse photocurrent $J_{\perp}$ as a function of the angle $\varphi$ defining the radiation helicity. Note that a small polarization independent offset $J_0$ is subtracted and the current $J_{\perp} - J_0$ is normalized to its maximum value. The data are obtained by applying oblique incidence radiation   with angle of incidence $\Theta = 33^{\circ}$.  (a) and (b) polarization dependencies of the photocurrent excited in samples \# A and  \# B by mid-infrared radiation with different photon energies $\hbar \omega$. Solid lines are fits to Eq.~(\ref{phenom1}), which corresponds to the polarization dependence of the sum of the CPGE and trigonal LPGE  derived in the theory presented in the main text, and described by Eq.~(\ref{phicirc}) and the first equation in Eq.~(\ref{ABDC}), respectively. The fit parameters used for the curves are: in the panel (a) - $J_C/J_L= 0.75$ and $\Phi =7^\circ$; in the panel (b) - $J_C/J_L= 3.2$ and $\Phi =6^\circ$. Drawings on top illustrate the polarization states at different angles $\varphi$. The inset in panel (a) shows the experimental geometry. The inset in panel (b) presents the magnitude of the circular photocurrent $J_{\rm CPGE} = (J(\sigma^+) - J(\sigma^-))/2$ measured for samples \#A and \#B. Note that in the inset the photoresponse to the mid-infrared radiation ($\hbar \omega =$ 100-130~meV) was obtained for intensities $I$ about 5~kW/cm$^2$, whereas in the THz range ($\hbar \omega =$ 4.4-13.7~meV) substantially higher intensities of the order of 100~kW/cm$^2$ were applied.
 	}
	\label{fig_02}
\end{figure}

The TI was grown as a bilayer heterostructure~\cite{Mayer2021} on SrTiO$_3$ (111) by means of molecular beam epitaxy (MBE). The heterostructures investigated here consist of one (sample \#A) or two (sample \#B and \#C) quintuple layers (QL) of Bi$_2$Se$_3$ as a seed layer and 10 (sample \#A) or 20~QL (sample \#B and \#C) {(Bi$_{1-x}$ Sb$_x$)$_2$(Te$_{1-y}$ Se$_y$)$_3$} (BSTS) layer, with $y \approx 0.85-0.90$ to maximize the band gap size~\cite{Ryu2016}, and $x \approx 0.7-0.75$ to compensate the $n$-type behavior of the BS seed layer. The $n$-type seed layer and $p$-type BSTS layer induce a band bending in growth direction as shown in Fig.~\ref{fig_01}. The heterostructure was capped in-situ by 7~nm aluminium oxide deposited by MBE. For photogalvanic measurements six Ti/Au-contact pads were placed with rectangular symmetry around the circumference of the $4 \times 5$~mm$^2$ sample by optical lithography. Fig.~\ref{fig_01}(b) shows a representative (here 5.6~QL BSTS) angle-resolved photoemission spectrum (ARPES), clearly displaying the Dirac point and the linear dispersion within the bulk bandgap \cite{Mayer2021}. Other ARPES images, supported by magnetotransport on gated samples, show that the position of the Fermi energy at the top surface was found to lie between the Dirac point (DP) and the conduction band minimum (CBM) for BSTS thicknesses larger than $\sim$5 QL~\ for the presented heterostructure concept and this particular BSTS composition~\cite{Mayer2021}.

\subsection{Methods}
\label{methods}

The photocurrents were studied by applying polarized laser radiation in the mid-infrared and terahertz range. The sources of mid-infrared radiation were two  different CO$_2$ laser systems: a medium power $Q$-switched laser with a pulse duration of 250~ns (repetition frequency~of~160~Hz)~\cite{Shalygin2006,Diehl2007} and a high power pulsed transversely excited atmospheric pressure (TEA) CO$_2$ laser providing 100~ns pulses with a repetition rate of 1~Hz~\cite{Ganichev2003,Ganichev2007}. These line-tunable lasers provide wavelengths in the range from  9.2 to 10.8~$\mu$m (32.6~THz$ \leq f \leq 27.8$~THz) corresponding to photon energies  $\hbar \omega $ from 135 to 115~meV.   The radiation was focused to a spot of about 1.5~mm diameter ($Q$-switch laser) and about 2~mm (pulsed laser), being much smaller than the sample size even at oblique incidence. This allowed us to avoid an illumination of the contacts or sample edges. The beam positions and profiles were checked with pyroelectric cameras~\cite{Ziemann2000,Drexler2012} or thermo-sensitive paper. Using different types of lasers allowed us to measure photocurrents in the radiation intensity range from $I = 0.2$ to 2~kW/cm$^2$ ($Q$-switch laser) and from 20 to 100~kW/cm$^2$ (TEA-laser). For measurements in the THz spectral range, we used a line-tunable  pulsed molecular laser  with NH$_3$  as the active medium operating at $\lambda =$ 90.5, 148, 280~$\mu$m ($\hbar \omega $ ranging from 13.7 to 4.4~meV)~\cite{Ganichev1982,Lechner2009}. The laser generated single pulses with a duration of about 100~ns and a repetition rate of 1~Hz, yielding a radiation intensity on the sample surface up to 800~kW/cm$^2$. The peak power of the radiation was monitored, depending on the system, with  photon-drag~\cite{Ganichev84p20}  detectors, mercury-cadmium-telluride (MCT) detectors~\cite{Rogalski2018} and pyroelectric power meters.

The geometry of the experiment is sketched in the inset of Fig.~\ref{fig_02}. The photocurrent was measured in unbiased structures via the voltage drop across a $600\: \Omega$ load resistor ($R_L$) in experiments with the pulsed laser, and across the
sample resistance in experiments applying  $Q$-switch laser.  The corresponding photosignal was recorded with a storage oscilloscope. Experiments were carried out at room temperature and down to 80~K, applying radiation at oblique and at normal incidence. In the measurements with oblique incident radiation, used to excite the circular photogalvanic effect,  the angle of incidence $\Theta_0$ was varied between $-40^\circ$ and $40^\circ$  ($\Theta_0 = 0$ corresponds to normal incidence) with the ($xz$) or ($yz$) plane of incidence, see the inset in Fig.~\ref{fig_02}. In our experiments, the  photoresponse was probed in directions perpendicular and parallel to the light incidence plane, i.e. transverse and  longitudinal arrangements, respectively, see insets in Fig.~\ref{fig_02} and~\ref{fig_03}.

The initial laser radiation polarization vector was oriented along the $y$-axis. To analyze the polarization dependencies of the photocurrent, we rotated $\lambda/2$- or $\lambda/4$- plates. In the former case, the radiation electric field vector $\bm E$ was rotated by the azimuth angle $\alpha$ with respect to the $y$-axis  while in the latter case, apart changes from the degree of linear polarization degree, we changed the radiation helicity $P_{circ}$  according to $P_{circ} = \sin 2\varphi$, where $\varphi$ is the azimuth of the quarter-wave plate. Consequently, for $\varphi = 45^\circ$ and 135$^\circ$ we obtained right-handed ($\sigma_+$) and left-handed ($\sigma_-$) circularly  polarized radiation with $P_{circ} = +1$ and -1, respectively.     
%
The polarization states for some angles $\varphi$ are illustrated on top of Fig.~\ref{fig_02}.

\section{Results}
\label{results}

\subsection{Experimental study of photogalvanic effects}

\begin{figure}[t]
	\centering
	\includegraphics[width=\linewidth]{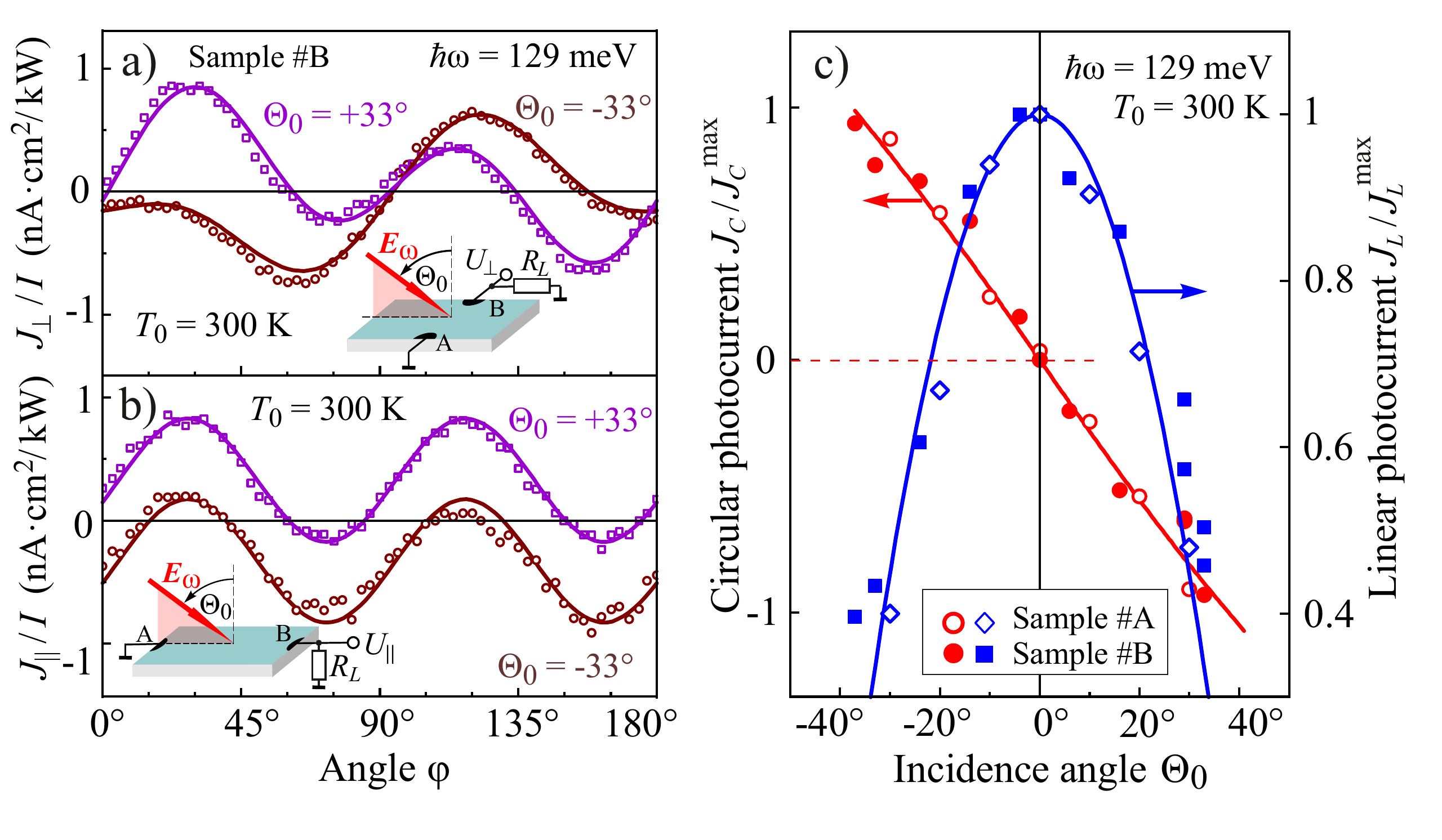}
	\caption{Transversal $J_{\perp}$, panel (a), and longitudinal $J_{||}$, panel (b), photocurrent as a function of the angle $\varphi$. The data are obtained for $\hbar \omega$ = 129 meV, angles of incidence $\Theta_0 = \pm 33^{\circ}$ and radiation intensities $I = 6.5$~kW/cm$^2$ (data for sample \#B) and 1.17~kW/cm$^2$ (data for sample \#A). In both arrangements the signal was picked up from the contacts AB, i.e., in the same crystallographic arrangement. 
		Corresponding experimental setups are shown in the insets in panels (a) and (b). 
		Solid lines in the panel (a) are fits after Eq.~(\ref{phenom1}), which corresponds to the sum of the CPGE and the trigonal LPGE described by Eq.~(\ref{phicirc}) and the first equation in Eq.~(\ref{ABDC}), respectively. The fit parameters used for the curves are: for $\Theta_0=33^\circ$: $J_C/J_L= 0.34/0.54$ and $\Theta_0=-33^\circ$: $J_C/J_L= -0.5/0.31$. 
		Solid lines in the panel (b) are fits after Eq.~(\ref{phenom2}), which corresponds to the the first equation in Eq.~(\ref{ABDC}). The fit parameter used for these curves is $J_L = 0.55$~nA.  Note that in the longitudinal geometry the CPGE vanishes. Angle $\Phi$ for all curves was $7^\circ$.
		Panel (c): Dependence of the CPGE and LPGE photocurrent contributions on the angle of incidence $\Theta_0$ for different samples. The photocurrent is normalized to its maximum values: $J_{\rm CPGE}^{\rm max}$ ($\Theta_0 = -40^\circ$) and $J_{\rm LPGE}^{\rm max}$ ($\Theta_0 = 0$). The data are obtained for $\hbar \omega$ = 129 meV, angles of incidence $\Theta_0 = \pm 33^{\circ}$ and radiation intensities $I = 6.5$~kW/cm$^2$ (data for sample \#B) and 1.17~kW/cm$^2$ (data for sample \#A). Solid lines are fits after Eqs.~(\ref{phicirc}) (CPGE) and ~(\ref{phitheory}) (LPGE). The fits are obtained assuming that the angle $\Theta$ for the radiation in the sample is equal to the angle of incidence $\Theta_0$.
	}
	\label{fig_03}
\end{figure}

Figures~\ref{fig_02} and \ref{fig_03}(a) show the polarization dependence of the transverse photocurrent $J_{\perp}$, excited by oblique incident mid-infrared radiation ($\Theta_0 = \pm 33^\circ$) and measured in the direction normal to the plane of incidence.  The  dependence, obtained for several photon energies ranging from 117 to 133~meV, in analogy to Ref.~\cite{Dyakonov2017,Plank2018,Olbrich2014}, is well described by
\begin{equation}
\label{phenom1}
J_{\perp} = J_{C} \sin 2\varphi + J_{L} \sin(4\varphi -3 \Phi
)/2 + J_0\:,
\end{equation}
%
where $J_{C}$,
$J_{L}$,
and $J_0$  are the magnitudes of the circular photocurrent,  linear contributions, and a polarization insensitive offset,  respectively. 
Note that the offset $J_0$ will be neglected in the remainder of the study since it only occured in a few of our measurements and was then found to be close to zero. The phase angle $\Phi$, discussed below in Sec.~~\ref{phenomtheory}, is 6$^\circ$ and 7$^\circ$ for samples \#A and \#B, respectively. 
%
%

Rotating the sample by 90$^\circ$ and measuring now the photocurrent generated in the direction parallel to the place of incidence (longitudinal photocurrent, $J_\parallel$), we observed that the circular component ($J_{C}$)  vanishes, and, in analogy to Ref.~\cite{Olbrich2014}, the current follows 
%
\begin{equation}
\label{phenom2}
J_{\parallel} = J_{L} \sin(4\varphi - 3 \Phi
)/2 + J_0^\prime\:,
\end{equation}
where 
$J^\prime_0$ is a polarization insensitive contribution. Figure~\ref{fig_03}(b)  shows the data obtained for sample \#B for $\hbar \omega =$ 129~meV and $\Theta_0 = \pm 33^\circ$. In contrast to the transversal photocurrent $J_{\perp}$, in the longitudinal one ($J_{\parallel}$)
the polarization insensitive contribution is comparable with $J_{L}$ and is odd in the angle of incidence.

Figure~\ref{fig_03}(c) presents the dependencies of the amplitudes of the circular and linear contributions on the angle of incidence $\Theta_0$. It demonstrates that the circular photocurrent is odd in the angle $\Theta_0$ and vanishes for normal incidence, whereas the linear one is even in the angle $\Theta_0$ and approaches its maximum at normal incidence.  The data can be well fitted by $J_C \propto \sin \Theta_0$ and $J_L \propto \cos \Theta_0$.

\begin{figure}[t]
	\centering
	\includegraphics[width=0.7\linewidth]{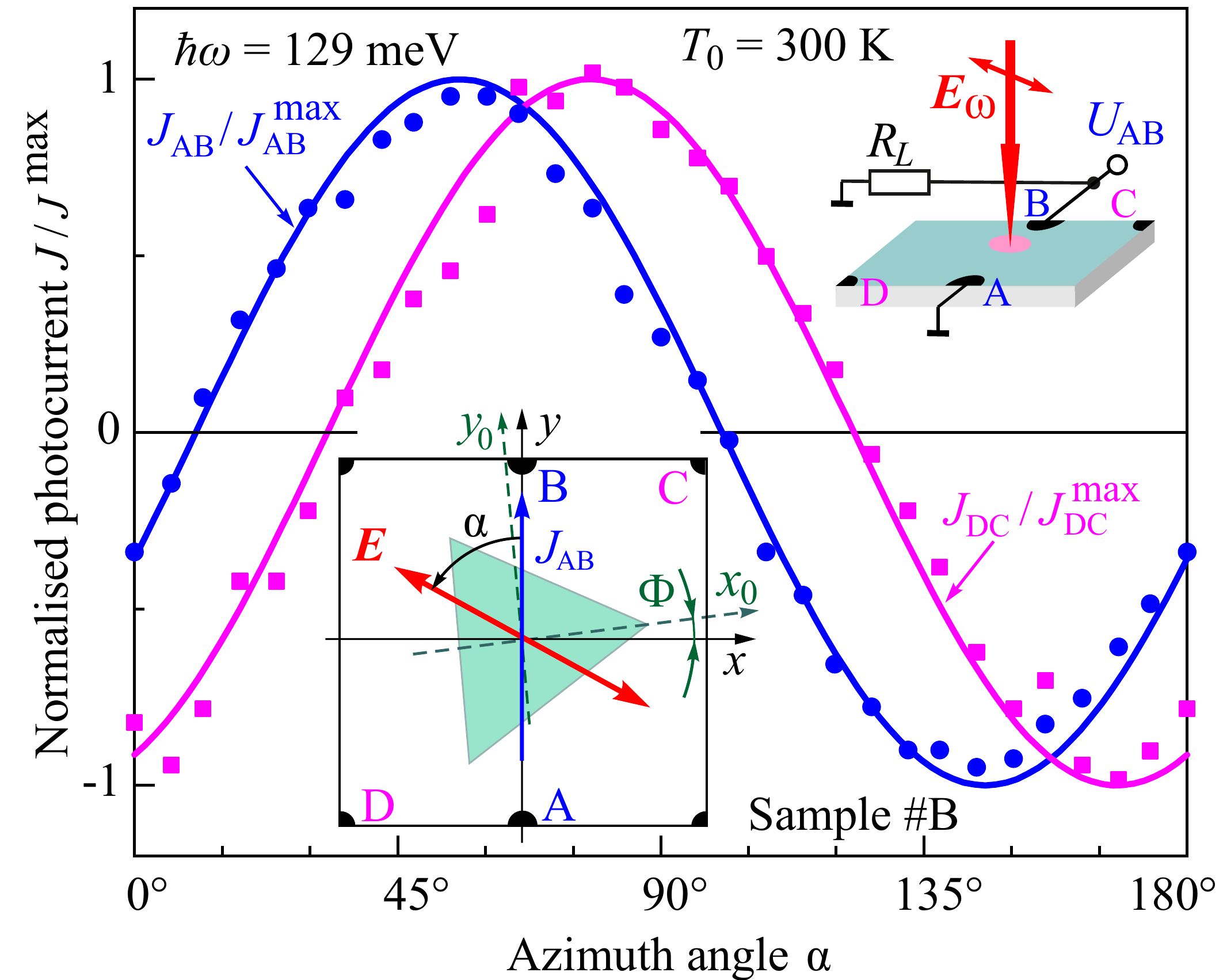}
	\caption{
		Dependence of the photocurrent normalized to its maximum value. The data are obtained applying normal incident linearly polarized radiation. The photocurrent is presented for the contact pairs AB and CD. Curves are fits after Eqs. (\ref{fit_alpha2}), which corresponds to the theoretical Eqs. (\ref{ABDC}) describing the trigonal PGE. Right inset shows the experimental setup. Left inset sketches the sample, contact pairs used for the measurement, and the ac electric field vector $E$ (double arrow). It also defines the azimuth angle $\alpha$ and presents schematically the orientation of axes along ($x_0$) and perpendicular ($y_0$) to one of the mirror reflection planes of the $C_{3v}$ point group illustrated as an equilateral triangle. Fits after Eqs. (\ref{ABDC}) yield that these axes are tilted with respect to the sample edges by $\Phi = 7^{\circ}$. The data are obtained for intensity $I = 1.17$~kW/cm$^2$. 
	}
	\label{fig_04}
\end{figure}

Our analysis presented below demonstrates that the  contribution to the photocurrent given by the coefficient $J_{L}$ is proportional to the degree of linear polarization. Consequently, it can be excited with linearly polarized radiation. This indeed has been observed.
Figure~\ref{fig_04} shows the results obtained in sample \#B for normal incident radiation and two in-plane directions (contact pairs AB and CD). The signals are well fitted by 
\begin{align}
\label{fit_alpha2}
J_{AB} = J_{L} \sin (2\alpha - 21^\circ), \nonumber \\
J_{DC} = J_{L} \sin (2\alpha- 21^\circ-45^\circ). 
\end{align}
As we show below, the overall behavior of the photocurrent  corresponds to that expected for the trigonal photogalvanic effect (terms proportional to the coefficients $J_{L}$) and to the circular photogalvanic effect (CPGE, the term proportional to the coefficient $J_C$).

\begin{figure}[t]
	\centering
	\includegraphics[width=\linewidth]{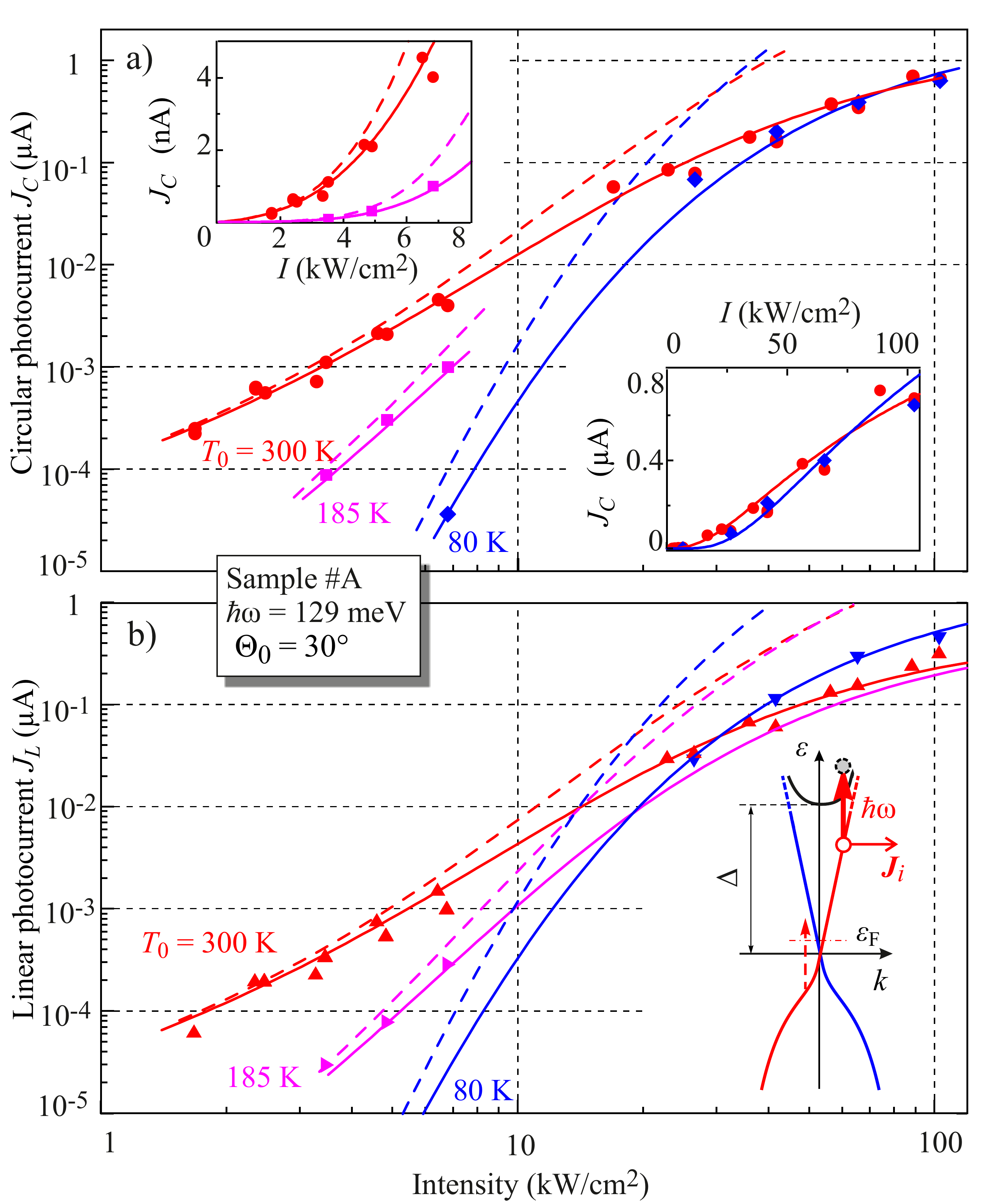}
	\caption{
Intensity dependencies of the CPGE (a) and LPGE (b) contributions to the photocurrent in sample \#A at different temperatures. The upper and bottom insets in panel (a) show the CPGE data in the double linear scale for the low and whole intensity ranges, respectively.
Dashed lines are fits after Eqs.~\eqref{superlinear} and ~\eqref{balance}, which disregard the photocurrent saturation. Solid lines are fits after Eqs.~\eqref{saturation}, ~\eqref{sublinear} and ~\eqref{balancesaturation}, which takes saturation processes into account.
The used fit parameters are:   $\varepsilon_i = 218$ meV, $\varepsilon_F = 8$  meV, $I_s=50$  kW/cm$^2$ , and $k_B \Delta T = 2 I$~[kW/cm$^2$].
%
The inset in panel (b) sketches the model of the photocurrent generation due to ``photoionization'' of the surface states (vertical solid arrow, $\hbar \omega = 130$~meV) in samples with the Fermi energy $\varepsilon_F$ close to the Dirac point. The spectra is extracted from the ARPES results, see Fig.~\ref{fig_01}(b). The horizontal arrow sketch the photocurrent contributions excited in the initial ($J_i$) state of the optical transitions. The dashed vertical arrow shows the direct optical inter-band transition. In the studied samples $\Delta = 270$~meV.
		}
	\label{fig_05}
\end{figure}

\begin{figure}[t]
	\centering
	\includegraphics[width=1.0\linewidth]{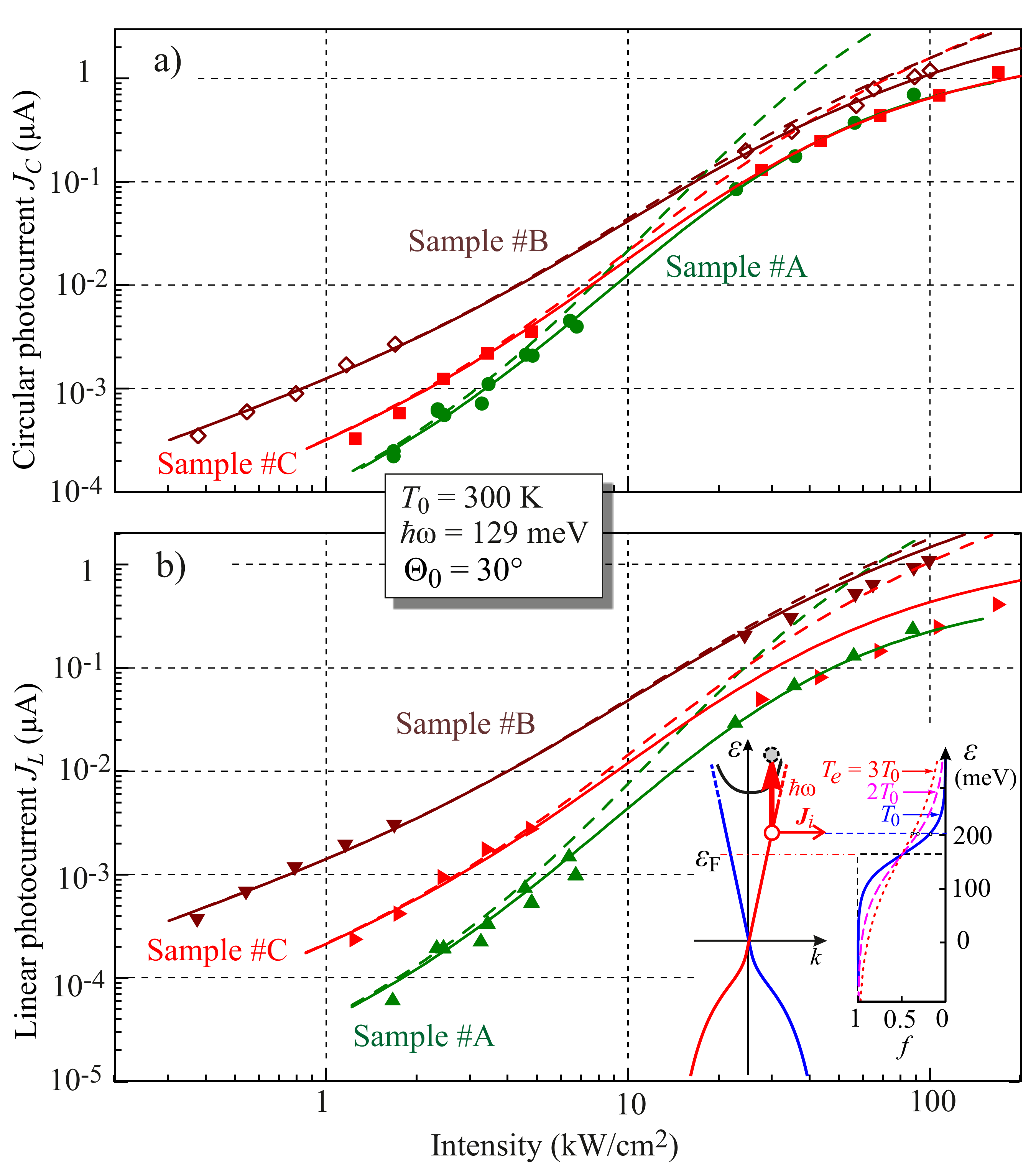}
	\caption{
		Intensity dependencies of the CPGE (a) and LPGE (b) contribution to the photocurrent in different samples at room temperatures. Dashed lines are fits after Eqs.~\eqref{superlinear} and ~\eqref{balance}, which disregard the photocurrent saturation. Solid lines are fits after Eqs.~\eqref{saturation}, ~\eqref{sublinear} and ~\eqref{balancesaturation}, which takes saturation processes into account.
The used fit parameters are:   $\varepsilon_i = 218$ meV, $\varepsilon_F =$ 8, 122 and 80 meV, $I_s=$ 50, 300 and 120  kW/cm$^2$ for samples \#A, \#B, \#C respectively , $k_B \Delta T = 2 I$. Note the substantial difference of  $\varepsilon_F$ used for fitting data in different samples. The inset in panel (b) shows the model of photocurrent generation for $\varepsilon_{\rm F}$ well inside the CB and its corresponding density of states for three different electron temperatures $T_e$.
}
	\label{fig_06}
\end{figure}


\subsection{Superlinear PGE intensity dependence}

Figure~\ref{fig_05} and \ref{fig_06} show that both the circular and linear PGE exhibit a strongly nonlinear intensity dependence: for small and moderate radiation intensities $I$ for increasing $I$ is seen a superlinear increase of the photocurrent magnitude, e.g. for room temperature data changing of $I$ by one order of magnitude results in the increase of the photocurrent by three orders of magnitude, see Figure~\ref{fig_05}. At high intensities the dependence becomes weaker and the photocurrent tends to saturate. These are surprising, and central observations, since according to phenomenological and microscopic theories photogalvanic currents are expected to scale linearly with the radiation intensity.  As we demonstrate in Sec.~\ref{absorption}, the observed nonlinearities can be well described by the intensity dependence of the radiation absorption. 
%

Comparing data for different samples, we observed that for samples \#B and\#C the magnitudes of the CPGE and LPGE photocurrents at low intensities are substantially larger than those detected in sample \#A, see Fig. \ref{fig_06}(a). At the same time the photoresponse nonlinearity detected in these samples, in particular in sample \#B, is significantly weaker.

Reducing the temperature from 300 to 185~K we observed that in sample \#A the signal at low intensities drastically reduces and the nonlinearity is more pronounced, whereas at high intensities the amplitudes of the circular and linear photocurrents and their intensity dependence stay almost the same for both temperatures see Fig. \ref{fig_05}. Similar results are obtained for $T=80$~K, see Fig. \ref{fig_05}.

All results described above were obtained applying mid-infrared radiation. Measurements with THz radiation revealed that at low frequencies the photocurrent vanishes, see the  inset in Fig.~\ref{fig_02}(b),
although an intensity one order of magnitude larger than in experiments with mid-infrared radiation of the TEA CO$_2$ laser was used.  This observation clearly demonstrates that both linear and circular PGE can not be caused by the Drude absorption, for which the signal substantially increases with the frequency decrease ($j \propto 1/(1+\omega^2\tau^2)$~\cite{Plank2018,Plank2018_2}, where $\tau$ is the momentum relaxation time). Note that the THz photon energies of our molecular laser used in this work are smaller by about an order of magnitude than that of the CO$_2$ laser. 
 Combining this argument together with the fact that the observed angle of incidence dependence can only be attributed to the excitation of the surface layer we conclude that the photocurrents are caused by direct optical transitions in the 3D TI topologically protected surface states, which are characterized by a linear energy dispersion, as shown in the ARPES data in Fig.~\ref{fig_01}. Note that the Rashba spin splitting, frequently discussed with respect to the surface states~\cite{Zhang2010,King2011}, can not solely be responsible for the photocurrent formation, because the PGE is caused by direct transitions and the used photon energies ($\approx 120 \divisionsymbol 140$~meV) are too large for transitions between Rashba split subbands.

Summarizing the experimental part, we have demonstrated that the illumination of BSTS samples with mid-infrared radiation results in linear and circular photogalvanic currents which are caused by direct optical transitions involving topological surface states. Importantly, the observed photocurrents are  characterized by a highly superlinear intensity dependence, which at high intensities tends to saturate.

\section{Phenomenological theory}
\label{phenomtheory}

The surface of the studied samples is described by the $C_{\rm 3v}$ point symmetry group. The phenomenological analysis yields that the PGE current density $\bm j$ is given by a sum of the circular ($\bm j^{\rm circ}$) and trigonal linear $(\bm j^{\rm tr})$ PGE contributions~\footnote{Note that, generally speaking besides these PGE currents, phenomenological theory yields for $C_{3v}$-symmetry systems several further photocurrents to which belong, LPGE excited at oblique incidence being odd in the angle of incidence~\cite{Weber2008} as well as linear and circular photon drag effects~\cite{Plank2016drag}. The analysis of the obtained results shows, however, that in the described experiments these photocurrents do not contribute substantially.}:
%
\begin{equation}
\label{total}
\bm j = \bm j^{\rm circ} + \bm j^{\rm tr}.
\end{equation}
The circular photocurrent is given by the following expression~\cite{Weber2008}
	\begin{equation}
	\label{j_CPGE}
	\bm j^\text{circ} = \gamma |\bm E|^2 \hat{\bm z}\times\bm \varkappa  = \gamma i(\bm E E_z^*-\bm E^*E_z).
	\end{equation}
Here $\gamma$ is the CPGE 
constant, $\bm E$ is the complex amplitude of the electric field acting on charge carriers, $\hat{\bm z}$ is the unit vector along the normal to the surface, and $\bm \varkappa$ is the photon angular momentum being maximal at circularly polarized radiation. 
It follows from Eq.~\eqref{j_CPGE} that the CPGE current is perpendicular to the incidence plane and it exists at oblique incidence only. For oblique incidence of elliptically-polarized radiation obtained using the quarter-wave plate, the CPGE current density is given by
	\begin{equation}
	\label{phicirc}
	j^\text{circ} = \gamma E_0^2 t_pt_s \sin{\Theta} \sin{2\varphi}.
	\end{equation}
Here $E_0$ is the incident light wave amplitude,
$\sin\Theta = \sin(\Theta_0)/n_\omega$ with $n_\omega$ being the refractive index, $t_p$ and $t_s$ are the Fresnel amplitude transmission coefficients, and $\varphi$ is the $\lambda/4$ plate rotation angle.  We used the relation  $P_\text{circ}=\sin{2\varphi}$ for the circular polarization degree. The CPGE resulting in opposite direction of the transversal photocurrent for right- ($\sigma^+$) and left- ($\sigma^-$) circularly polarized radiation is clearly detected in the experiment, see Figs.~\ref{fig_02} and~\ref{fig_03}(a). Experiments also confirm the expected absence of the CPGE for the longitudinal photocurrent, see Eq.~(\ref{j_CPGE})  and Fig.~\ref{fig_03}(b).
Equation~\eqref{phicirc} shows that the angle of incidence dependence of the circular photocurrent is described by $j^\text{circ} \propto  \sin{\Theta_0}$. This dependence describes well the experiment, see Fig.~\ref{fig_03}(c).

In contrast to the invariant form of the CPGE current Eq.~\eqref{j_CPGE}, the trigonal LPGE current 
$\bm j^{\rm tr}$ is determined by the orientation of  the electric vector in respect to crystallographic axes $(x_0,y_0)$ along and perpendicular to one of the mirror reflection planes of the C$_{3 {\rm v}}$ point group, see the inset in Fig.~\ref{fig_04}. The current density of the trigonal LPGE is given by~\cite{Olbrich2014}
	\begin{equation}
	\label{j_trig}
	j_{x_0}^{\rm tr} + ij_{y_0}^{\rm tr}= \chi (E_{x_0}-iE_{y_0})^2,
	\end{equation}
	%
where $\chi$ is the trigonal LPGE constant. The current and electric field components in the coordinate system $(xy)$ used in the experiments are related as follows
	\begin{equation}
		\label{j_trig_xy}
	j_x^{\rm tr}+ij_y^{\rm tr} = \chi (E_{x}-iE_{y})^2\text{e}^{3i\Phi},
	\end{equation} 
where $\Phi$ is an angle between the axes $x$ and $x_0$.

The above phenomenological expressions \eqref{total}-\eqref{j_trig_xy}  demonstrate that the photocurrent perpendicular to the incidence plane is a sum of two contributions at oblique incidence: $\bm j_\perp = \bm j^{\rm tr}_\perp + \bm j^{\rm circ}$, while the photocurrent flowing in the incidence plane is due to trigonal LPGE only, $\bm j_\parallel = \bm j^{\rm tr}_\parallel$, and it is maximal at normal incidence.

The trigonal LPGE is given by the in-plane components of the radiation electric field and, therefore, is proportional to the degree of linear polarization. For the photocurrent  measured along AB and DC lines  in Fig.~\ref{fig_04} we have
	\begin{align}
	\label{ABDC}
	j_{AB}  = \chi |\bm E|^2\sin{(2\alpha-3\Phi)}, \nonumber \\
	j_{DC} = 
	\chi |\bm E|^2\sin{(2\alpha-3\Phi-45^\circ)}.
	\end{align}
Here $\alpha$ is an angle between the linear polarization axis and the AB line. These expressions explain the $\pi$-periodic dependence of the currents $J_{AB}$ and $J_{DC}$ shown in Fig.~\ref{fig_04} and fitted by Eqs.~\eqref{fit_alpha2}. The observed phase shift shows that, in the studied sample, the $x_0$- and $y_0$-directions are rotated by the angle $\Phi = 7^\circ$ in respect to the sample edges. 

For oblique incidence of elliptically-polarized light obtained by using the $\lambda/4$ plate, see Figs.~\ref{fig_02} and~\ref{fig_03},  the trigonal LPGE current is given by
	\begin{multline}
	\label{phitheory}
	j_\parallel^{\rm tr}+ij_\perp^{\rm tr}
	= - {\chi E_0^2\over 2} \text{e}^{3i\Phi}\left({3t_s^2-t_p^2\cos^2{\Theta}\over 2} \right. \\
\left.	+ {t_s^2+t_p^2\cos^2{\Theta}\over 2}\cos{4\varphi} 
	- i t_s t_p\cos{\Theta}\sin{4\varphi} \right).
	\end{multline}
Here we assumed that at $\varphi=0$ the radiation is $s$-polarized. For both components of the trigonal LPGE current we obtain
%
	\begin{equation}
		\label{j_tr_phi}
	j_{\perp,\parallel}^{\rm tr} =
	\pm{\chi E_0^2\over 2} [A\sin{(4\varphi-\varphi_0)} - B],
	\end{equation}
where $\varphi_0$, $A$ and $B$ are the phase,
the current amplitude and the background signal, respectively. They are given by
	\begin{multline}
	\tan{\varphi_0}={t_s^2+t_p^2\cos^2{\Theta}\over 2t_pt_s\cos{\Theta}}\tan{3\Phi}, \\
	A= \sqrt{(t_pt_s\cos{\Theta})^2\cos^2{3\Phi}+(t_s^2+t_p^2\cos^2{\Theta})^2\sin^2{3\Phi}/4},
	\\
	B = {3t_s^2-t_p^2\cos^2{\Theta}\over 2}\sin{3\Phi}.
	\end{multline}
%
%
Note that for $\Theta_0 \leq 40^\circ$ we obtain from the first equation the angle $\varphi_0 \approx 3\Phi$. 
%
%
For small angles $\Phi$ at which $\sin{3\Phi} \approx 0$ and considering $n_\omega = 1$ we obtain $A \propto \cos{\Theta_0}$. At normal incidence $t_p = t_s$ and for circularly polarized radiation ($\varphi = 45^\circ$ and 135$^\circ$) the first and the second terms in the rectangular brackets in Eq.~(\ref{j_tr_phi}) becomes equal to each other but have opposite signs, i.e., as expected the LPGE vanishes.
	
The $\pi/2$-periodic dependence of the longitudinal photocurrent presented in Fig.~\ref{fig_03}(b) is perfectly described by the dependence $j_\parallel^{\rm tr}(\varphi)$, see	Eq.~\eqref{j_tr_phi}.  The perpendicular to the incidence plane photocurrent  $\varphi$-dependence shown in Fig.~\ref{fig_03}(a) and fitted by Eq.~\eqref{phenom1} is exactly given by the sum $j_\perp^{\rm tr}(\varphi)+j^{\rm circ}(\varphi)$. The incidence angle dependence of the trigonal LPGE also describes the experimental results demonstrated in Fig.~\ref{fig_03}(c). 
%
We emphasize, that in experiment the angular dependencies of the LPGE $j_{\rm LPGE} \propto \cos{\Theta_0}$ is defined by the angle of incidence $\Theta_0$ and not by the angle $\Theta$,  see Fig.~\ref{fig_03}(b). The trigonal LPGE effects are defined by the electric field components acting on charge carriers in material: in-plane $E$-fields for the trigonal LPGE and a product of in-plane and out-of plane component for the CPGE, see Eqs.~\eqref{j_CPGE} and~\eqref{j_trig_xy}. Consequently, in materials with large refractive indices, at oblique incidence as large as $\Theta_0 = \pm 40^\circ$ the in-plane component should remain almost unchanged, whereas the out of plane component should be close to zero. The fact that both photocurrents vary in the experiment with the angle of incidence shows that the photocurrent is excited in the thin surface layer for which the refraction  is not yet formed.


Summarizing, we demonstrate that the detected dependencies of the PGE currents are in full agreement with the phenomenological theory.

\section{Theory of mid-infrared radiation absorption}
\label{absorption}
\subsection{Spectral dependence of the absorbance}
\label{absorbanceJan}

\begin{figure}[t]
	\centering
	\includegraphics[width=\linewidth]{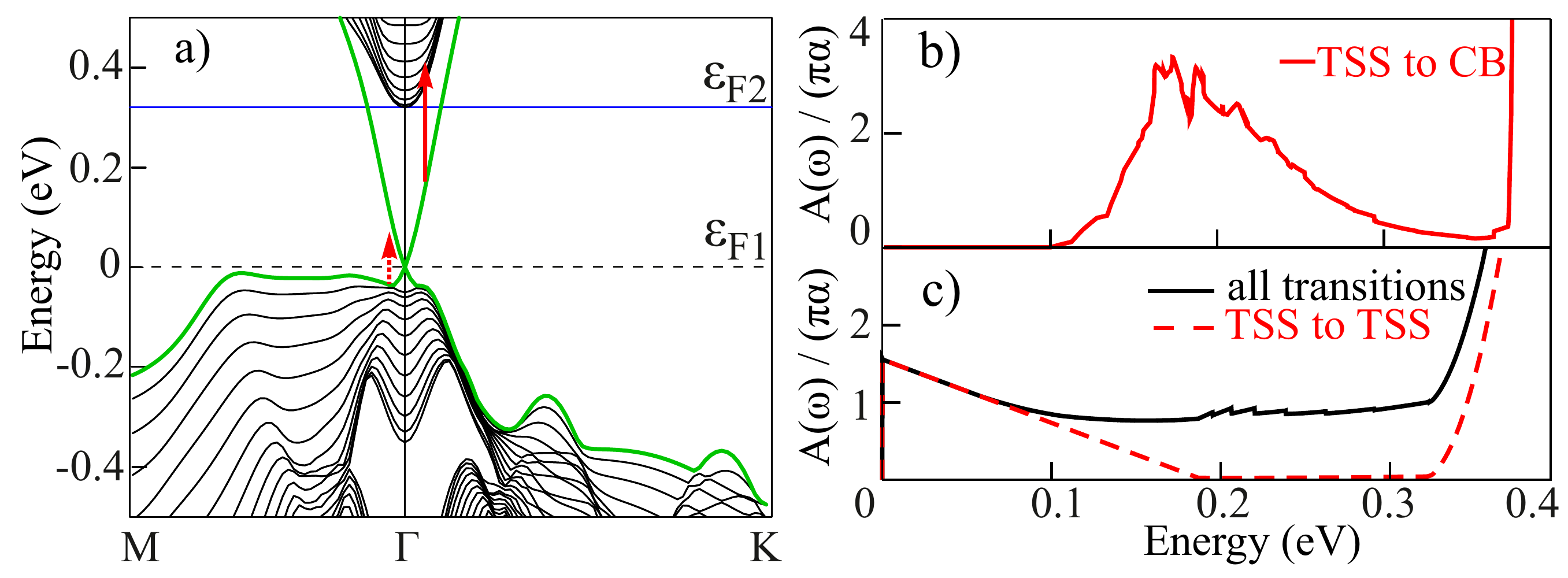}
	\caption{ Panel a: Electronic structure from 13QL of Bi$_2$Se$_3$ slab calculations,
		with topological surface state in green and bulk states in
		black. Panel b:  Low energy optical absorbance
		$\sigma_{1xx}(\hbar\omega)$ between the topological surface
		state TSS and conduction band with respect to the shifted Fermi energy
		$\varepsilon_{\rm F2}$ (blue line) at 282meV. Panel c: Total (black
		line) absorbance and the contribution including only
		direct optical transitions (dashed line) from TSS to the TSS
		with  as-calculated  Fermi energy $\varepsilon_{\rm F1}$ located at the
		Dirac point. 
}
	\label{fig_08}
\end{figure}

To estimate the value of optical absorbance for two relevant channels,
see inset of Fig.\ref{fig_05}b), we performed band structure
calculations with the local density approximation of
Bi$_2$Se$_3$ using the linear muffin tin orbital (LMTO) method
\cite{Andersen1975}. We used the fully relativistic PY LMTO
computer code \cite{Antonov2004}  where relativistic effects are treated on the level of spin
density functional theory by solving corresponding Dirac Hamiltonian.
The self consistent calculations of the surface has been done for
the slab geometry considering 13 quintuple layers of
Bi$_2$Se$_3$ using experimental lattice parameters.
This thickness is sufficient to minimize
hybridization between two TSSs and it leads to the negligible gap
($~10^{-3}$~meV) in the DP. The $xx$ component of the absorptive part of the optical
conductivity tensor for finite frequencies has been determined  by
means of the Kubo–Greenwood linear-response formalism
\cite{Antonov1999,Ebert1996}. In order to resolve inter band transitions
between TSS and conduction band we used tetrahedron method on a
very dense 64x64x16 k-mesh. The calculated optical conductivity
has been normalised to the slab thickness of $14.318$~nm to obtain
absorbance. 

In the Fig.~\ref{fig_08}a) we show
calculated band structure along M$\Gamma$K path in the Brillouin
zone with highlighted TSS (green line). The calculated Fermi
energy $\varepsilon_F1$ (dashed line) is always at the DP. 
The energy  denoted as $\varepsilon_F2$ (blue line) has been used
to calculate the  absorbance from transitions between the TSS and conduction
band (CB). In the Fig.~\ref{fig_08}b) photon energy dependence of
absorbance including all dipole allowed transitions between TSS
and CB is showed. The maximum absorbance of about 3 in units of
$\pi$ times the fine-structure constant $\alpha$ at
$\hbar\omega =$ 170~meV 
which is about three times larger then direct transitions between
TSSs as shown in Fig.~\ref{fig_08}c). It is worth to note that
the value of photon energy offset (100~meV in Fig.~\ref{fig_08}b))
very sensitively depends on the details of the band dispersion. In
particular, the finite size slab calculations limits the k-width of
the CB. Therefore the calculated absorbance maximum at 170~meV is
in good agreement with $\hbar\omega$ of 129~meV used here. Finally,
in Fig.~\ref{fig_08}c) the absorbance including  transitions
between TSS and TSS (dashed line) as well from valence band to the
TSS (full line). For the two dimensional materials like
e.g. graphene, with its characteristic linear $p_z$ orbital based
linear dispersion, the constant infrared absorbance of 1
$\pi\alpha$ has been measured \cite{Nair2008}. However, as it
can be observed from the
Fig.~\ref{fig_08}c) in the case of TSS of in Bi$_2$Se$_3$ we predict
linearly decreasing absorbance at low photon energies. This behavior
is due to the non-linear dispersion of the TSS close to the valence
band and it is connected  to the strong layer
dependence of the orbital p$_z$ and p$_{x,y}$ character of the
TSS. This is reflected by the warping of the constant energy
surface. \cite{Zhu2013,Sanchez-Barriga2014}.

Summarizing, in this section we showed that the infrared absorbance related to
the optical transition between TSS and CB  are dominant as compared to
the direct transition between TSS.

\subsection{Optical transitions between surface and bulk states in 3D~TI}
\label{absorbanceLG}

We consider the process of ``photoionization'' of the surface states in 3D TI where the carrier makes a transition from the surface DF state to the bulk states. 
%
The bulk 
states 
%
%
dispersion in the conduction and valence bands can be described by~\cite{Zhang2009,Liu2010}
\begin{equation}
E_{c,v}(\bm k)= \pm \sqrt{\Delta^2 + (A_1k_z)^2 + (A_2 k_\parallel)^2},
\end{equation}
where $\bm k_\parallel$ is a projection of the wavevector on the surface, $k_z$ is the component normal to the surface, $A_{1,2}/\hbar$ are the anisotropic Dirac velocities, and $\Delta$ is a half of the bulk bandgap, see inset in Fig.~\ref{fig_05}~(b).

The 
surface states
are localized near the surface $z=0$ being plane waves in the surface plane characterized by the in-plane wavevector $\bm k_\parallel$. They have the conical dispersion 
$E_{s,\pm} = \pm A_2 k_\parallel$. 

For the photoionization process, the energy conservation law yields
\begin{equation}
\hbar \omega + A_2 k_\parallel = E_c(k_\parallel, k_z),
\end{equation}
which yields for the initial energy $\varepsilon_i = A_2 k_\parallel$ 
\begin{equation}
\varepsilon_i = {\Delta^2 - (\hbar\omega)^2 + (A_1k_z)^2 \over 2\hbar\omega}.
\end{equation}
Here $k_z$ is the projection of the wavevector of the final bulk state on the normal to the surface. The characteristic values of $k_z$ are in the range $0 \leq |k_z| \lesssim 1/l$, where $l=\Delta/A_1$ is the surface state localization length. Therefore we have an estimate for the energy of the initial  state
\begin{equation}
{\Delta^2 - (\hbar\omega)^2  \over 2\hbar\omega} \leq \varepsilon_i \lesssim {2\Delta^2 - (\hbar\omega)^2  \over 2\hbar\omega} .
\end{equation}
For $\Delta = 270$~meV and $\hbar\omega =129$~meV we obtain the range 218~meV $\leq \varepsilon_i \lesssim$ 500~meV.

%


\section{Discussion}

\subsection{Introductory Notes}

The analysis of our data in the framework of the developed phenomenological theory demonstrates that the observed photocurrents are caused by the circular and trigonal linear photogalvanic effects excited in the two-dimensional surface states, see Sec.~\ref{phenomtheory}. Furthermore, from the spectral behavior, showing that the photocurrent 
%
%
is observed at high frequencies and vanishes as the frequency decreases, see Fig.~\ref{fig_02}, we can conclude that both CPGE and LPGE are  caused by direct optical transitions, 
whereas the mechanisms related to indirect Drude-like transitions in the discussed experiments play a negligible role. In the mid-infrared range the direct optical transitions can be caused by two mechanisms: inter-band transitions between the DF states or their ``photoionization'' when one of the final or initial states lies in the three-dimensional conduction or valence band, respectively, see Sec.~\ref{absorbanceJan}.  

Both possible channels of the radiation absorption may cause the circular and linear PGE currents. 
The trigonal photogalvanic effect has been considered for the low frequency range, where it is shown to be caused by the asymmetric scattering of carriers driven by the radiation electric field~\cite{Olbrich2014}. In the experiments described above we used, by contrast, a high frequency radiation and the photocurrent is formed at direct optical transitions. Such photocurrents are always a sum of two contributions, the shift and the ballistic ones~\cite{Sturman1992,Ivchenko2005,Weber2008,GolubIvchenko_JETP_2011,Kim2017,Sturman2020shift}. While the former contribution is due to shifts of the electron wavepackets in real space, occurring in the process of the optical absorption, the latter is caused by the momentum scattering following the direct optical transition.
The theory of the circular PGE in Dirac fermion systems has been developed for transitions from the surface states to continuum considering HgTe 2D and 3D TI~\cite{Artemenko2013,Dantscher2017,Pan2017,Durnev2019} and for the interband transitions considering TIs~\cite{Hosur2011,Junck2013,Entin2016}, graphene-based systems~\cite{Kiselev2011,Ivchenko2012,Glazov2014,Hipolito2016,Candussio2020}, and Weyl semimetals~\cite{Juan2017,Chan2017,Ma2017,Golub2018,Leppenen2019,Ji2019,Chang2020,Rees2020}.  These results can be straightforwardly extended to the  material under study and will not be discussed here. 

While the trigonal LPGE photocurrent has been studied in a regime where it depends linearly on the light intensity~\cite{Olbrich2014,Plank2018}, the nonlinear regime has not been addressed so far.
Our experiments reveal that the PGE currents excited in the surface states may exhibit a highly nonlinear behavior already at moderate intensities. The  photocurrent saturation observed at very high intensities is not surprising. Such a behavior of the photogalvanic current has previously been detected in different semiconductor systems and even applied e.g. to determine spin relaxation times in III-V quantum wells~\cite{Ganichev2002,Schneider2004}. The superlinearity detected at rather low intensities, however, has not been reported yet. Below, we show that both the superlinearity and the photocurrent saturation are caused by the intensity dependence of the radiation absorption. 
%
		Our analysis reveals that in all samples the CPGE and LPGE are caused by the photoionization of the surface states.

\subsection{Analysis of the observed deviations from linearity in intensity}

We begin with the analysis of the superlinearity observed in all samples excited by moderate intensities below 10 kW/cm$^2$, see e.g the left top inset in Fig.~\ref{fig_05} presenting in double linear scale plot the intensity dependence of the CPGE excited in sample \#A at room temperature as well as Figs.~\ref{fig_05} and \ref{fig_06} where the data for all samples are presented in double-logarithmic presentation. While such a dependence may in principle be described in terms of two-photon absorption, our estimation shows that in the mid infrared range such transitions will yield considerable contributions only for intensities by many orders of magnitude larger as those used in our work~\cite{Candussio2021}. At the same time, the observed superlinearity can be explained for any level of the radiation intensities by considering the redistribution of carriers in the energy space which results from the radiation-induced electron gas heating. The insets in Figs.~\ref{fig_05}~(b) and \ref{fig_06}(b) schematically shows the generation of the PGE photocurrent caused by the ''photoionization'' of the surface states. The photocurrent caused by direct optical transitions is proportional to the radiation absorption and, consequently, to the difference of the initial and final  states occupancy of the direct transition  $[f_0\qty({\varepsilon_i})-f_0\qty({\varepsilon_f})]$, where $\varepsilon_i$ and $\varepsilon_f$ are the energies of these states. In the discussed experiments the photon energies $\hbar \omega \sim 100$~meV are much higher than the thermal energy $k_B T_0$ (26~meV for experiments at room temperature and lower for low temperature measurements). Thus, we can consider that the final states are unoccupied, which simplifies the consideration. 

We begin with the results obtained on sample \#A for which the position of the Fermi level $\varepsilon_F \sim 0$ is inferred from gated magneto-transport measurements~\cite{Mayer2021}. In this case the initial state of the ''photoionization''  transition, see inset in Fig.~\ref{fig_05}~(b) , has the energy higher than the Fermi level. Thus, increase of the electron temperature would result in the increase of the initial state population and, consequently in the increase of the photocurrent amplitude given by
\begin{equation}
\label{superlinear}
j_{\rm PGE} = \frac{C I \eta_0}{1+\exp\qty[\frac{\varepsilon_i-\varepsilon_F}{k_B(T_0+\Delta T)}]} \, ,
\end{equation}
where $\Delta T = T_e-T_0$ is intensity dependent increase of the electron temperature $T_e$  in respect to the lattice temperature~\footnote{We assume that the electron-electron collision time is much shorter than the energy relaxation time. Under this condition, the electron gas establishes a temperature $T_e$, which is different from the temperature of the lattice.}
and $C$ is a prefactor. 
The magnitude of the electron temperature $T_e$ is determined by the competition between power absorption and energy loss, and can be obtained from the energy balance equation~\cite{Ganichev2005}
 %
 \begin{equation}
 \label{balance}
\eta_0 I = n_s {k_\text{B}\Delta T\over \tau_\varepsilon} \, .
 \end{equation}
Here $\eta_0$ is the total absorbance involving all possible absorbance channels, considered in Sec.~\ref{absorbanceJan}, $n_s$ is the electron surface density, and $\tau_\varepsilon$ is the  energy relaxation time of electron gas.

Combining Eqs.~\eqref{superlinear} and ~\eqref{balance} we obtain the intensity dependence of the PGE current. Note that in this model the variation of the photocurrent magnitude with the radiation intensity is expected to be the same for both CPGE and LPGE. However, the prefactors and their variation upon change of temperature are different for these two mechanisms and reflect their microscopic details.  Dashed lines in Figs.~\ref{fig_05}(a) and (b) show that for low intensities ($I < 10$~kW/cm$^2$) the experimental dependencies of both CPGE and LPGE currents can be well fitted by the above equations. The curves obtained for different lattice temperatures show that in agreement with the above equations a decrease of the lattice temperature results in a higher nonlinearity. Importantly all fits were obtained for fixed values of the Fermi level $\varepsilon_F = 8 $~meV and the energy of initial states $\varepsilon_i =218$~meV. The latter was calculated in Sec.~\ref{absorbanceLG}. The best fits were obtained for $k_\text{B}\Delta T~ {\rm [meV]} = 2 I$, where radiation intensities are given in kW/cm$^2$. Consequently, for the room temperature data set and the largest intensity of the Q-switch laser $I \approx 8$~kW/cm$^2$ (data points for $I < 10$~kW/cm$^2$) we obtain that electron temperature $T_e \sim 1.6 T_0$.   
%

The observed superlinear intensity dependence of both photocurrents indicates that the main mechanism of their formation is based on the direct transitions from the surface sates to continuum, whereas the direct inter-band optical transitions does not yield a measurable contribution.  Indeed, in this case the initial state of the optical transition ($\varepsilon_i = - \hbar \omega /2$) is placed below the Fermi energy and final one ($\varepsilon_f = \hbar \omega /2$) above it. Thus the electron gas heating would reduce the difference between the states occupation $[f_0\qty({\varepsilon_i})-f_0\qty({\varepsilon_f})]$, which should result in the sublinear intensity dependence, which contradicts with experiment. 

Equations~\eqref{superlinear} and~\eqref{balance} also describe well the results obtained in samples \#B and \#C at room temperature. These samples are designed and grown in a way that the Fermi levels lie closer to the CBM, above the Dirac point~\cite{Mayer2021}, see the inset in Fig.~\ref{fig_06}(b). The increase of the Fermi energy results in the increase of the population of the initial state, and, consequently, should lead to larger magnitudes of both photocurrents. This is indeed observed in the experiment, see Fig.~\ref{fig_06}. The best fits of the data in a low intensity range ($I < 10$~kW/cm$^2$) were obtained with the Fermi energies $\varepsilon_F =122$~meV for sample \#B and 80~meV for sample \#C. Also in these samples a raise of the electron temperature leads to an increase of the initial state population, see right part of the inset in Fig.~\ref{fig_06}(b),  and, consequently to the superlinear behavior with rising radiation intensity. 

While Eqs.~\eqref{superlinear} and~\eqref{balance} describe well the observed superlinear behavior at low intensities ($I < 10$~kW/cm$^2$) they, in this form, do not explain the  observed saturation at high intensities. Saturation of absorbance caused by direct optical transitions is in fact not surprising. Such a process caused by slow relaxation of photoexcited carriers has been previously detected in a great variety of semiconductor systems,
for textbooks see e.g.~\cite{Ganichev2005,Saleh2019}, including BiTe-based 3D TIs, for review see e.g.~\cite{Autere2018}. 
Assuming that the absorbance saturates at high intensities as
 \begin{equation}
\label{saturation}
\eta(I)  = \frac{\eta_0}{1+ I/I_s}\, ,
\end{equation}
%
where $\eta_0$ is low power absorbance  and $I_s$ is the saturation intensity we re-write Eq.~\eqref{superlinear} in form
\begin{equation}
\label{sublinear}
j_{\rm PGE} = \frac{C \eta(I) I}{1+\exp\qty[\frac{\varepsilon_i-\varepsilon_F}{k_B(T_0+\Delta T)}]} \, ,
\end{equation}
with $\Delta T$ defined from the balance equation applying $\eta(I)$
\begin{equation}
\label{balancesaturation}
\eta(I) I = n_s {k_\text{B}\Delta T\over \tau_\varepsilon} \, .
\end{equation}
Corresponding fits, which are shown by solid lines in Figs.~\ref{fig_05} and \ref{fig_06}, describe well the data in the whole range of the studied intensities, with the saturation intensities $I_s=50 - 300$~kW/cm$^2$. In general, the saturation intensity is defined by the reciprocal energy relaxation time and the absorption cross-section, see e.g.~Refs.~\cite{Ganichev2005,Saleh2019}. A detailed study of the energy relaxation time need further experiments and is a subject of an independent research.

\section{Conclusions}

We demonstrate that the observed PGE result from direct transitions from the top topologically protected surface state to bulk conduction band. Investigating the intensity-dependence of the PGE we detected a strong superlinear behavior at low and moderate radiation intensities and a sublinear one (saturation) at high intensities. The superlinear behavior is demonstrated to be a consequence of the radiation induced electron gas heating, whereas the saturation results from slow energy relaxation of the  photoexcited carriers. Our analysis of the photogalvanic effects nicely illustrates the versatility and degree of control provided by the studied bilayer BSTS/BS heterostructures, both of which were inferred earlier from magneto-transport experiments on gated samples~\cite{Mayer2021}. Indeed, here, we conclude that the heterostructures provide topologically protected top surface states for which the Fermi energy lies between the Dirac point and the bulk conduction band minimum. The position of the Fermi energy is quite precisely controlled via the thickness of the BS and BSTS layers. As conjectured from this bilayer heterostructure concept~\cite{Mayer2021}, in sample \#A the Fermi energy lies very close to the Dirac point of the surface state in the as-grown heterostructure, while in samples \#B and \#C it is found to be moved up further towards the conduction band minimum. These conclusions, at the same time, illustrate how the analysis of deviations from the linear dependence in intensity of the linear and circular photogalvanic effects in 3D topological insulators provides a quantitative room temperature tool to access important materials parameters such as the position of the Fermi energy at the surface state or the energy relaxation times.

%

\section{Acknowledgments}
The support from from the Deutsche Forschungsgemeinschaft (DFG, German Research Foundation) – Project-ID 314695032 – SFB 1277  (project A01 and A04), the Elite Network of Bavaria (K-NW-2013-247), and the Volkswagen Stiftung Program is gratefully acknowledged. L.E.G. thanks the financial support of the Russian Science Foundation (Project 20-12-00147) and the Foundation for the Advancement of Theoretical Physics and Mathematics ``BASIS''. S.D.G. acknowledge the support of the the IRAP program of the Foundation for Polish Science (grant MAB/2018/9, project CENTERA). J. M. thank the CEDAMNF Project financed by the Ministry of Education, Youth and Sports of Czech Republic, Project No. CZ.02.1.01/0.0/0.0/15\_003/0000358 and the Czech Science Foundation (GACR),Project No. 20-18725S. We thank J. Fujii and I.Vobornik from the APE-LE beamline at the Elettra synchrotron inTrieste, Italy, for helping with the ARPES measurements.


\bibliography{references_f2}

\end{document}